# NEW ELEMENTS

# for a

# NETWORK (including BRAIN) GENERAL THEORY

# during

# LEARNING PERIOD

**FOR A BETTER UNDERSTANDING OF ARTIFICIAL INTELLIGENCE EFFICIENCY ?**

**Version N° 5** **A relevant progress?**

**By Jean PINIELLO**

JULY 2018

## INTRODUCTION

The purpose of this study is to understand the evolution of the so called intelligent **networks** (insect societies without leader, cells of an organism, **brain**, artificial neural networks,….), and more precisely during their learning period.

No doubt that researches currently implemented to explore the **brain**,as *Human Brain Project*, will allow us to visualize each neuron during its evolution and to find new therapeutics which will improve very significantly the health of many people, for instance by curing neurodegenerative deseases. And this will be a remarkable progress.

Similarly, **Artificial Intelligence** will make easier many aspects of our life (autonomous vehicles, diagnosing deseases, games,….), and we can also notice the recent discovery of the **trees intelligence**, which allows them to communicate among themselves , mainly to face predators.

All these examples have a common characteristic: **they use intelligent networks**. It proves that intelligence can emerge when elements exchanging information are connected by a network structure.

But the problem is that, in spite of using big data, statistic methods, deep learning, etc….,
**one does not know how this intelligence emerges and cannot explain scientifically, by a predictive mathematical model of the network, how the latter evolves.**

Generally, people who study this problem try to find a model of an element (the neuron for instance) so as to get to a model of the whole network (approach bottom/up). But we think this approach cannot give knowledge of the global network behaviour: indeed, even if we suppose we have perfectly modelized a neurone and that we connect all the neurons together, we probably shall find remarkable evolution of the whole structure (the network); **but we will be incapable to explain scientifically, mathematically, why and how these results occur.**

On the contrary, we think that it is better to consider directly the **whole network** to find a mathematical model of it (approach top/down).

So we do a **first hypothesis** that a general theory, **common to all networks**, can be developped, which would allow us to simulate mathematically the behaviour of any intelligent network (and particularly the **brain**).

Our **second hypothesis** is that this **network theory can be expressed by a formalism inspired by the one of *Quantum mechanics***, and more precisely the one of ***Quantum field theory*** (principle of stationary action, gauge fields, invariance by symmetry transformations,….), and may be also by the one of ***Quantum gravity***.

We have yet published a **Version N°2** of a study called

***"Elements for a network general theory during apprenticeship period"***

which can be consulted on the website
www.theorie-reseau-piniello.fr

or on the website of archives **HAL** www.hal.archives-ouvertes.fr/hal-00713298

or on the website **arXiv** www.arxiv.org/abs/1301.2959v2

But this Version N°2 has the inconvenient to describe the behaviour of a network by the mean of the evolution equations of **each element** of it; taking into account the considerable number of elements which can contain a network (for example the brain), this does not seem very realistic!....

So we publish the present **Version N°5** which, may be, offers a relevant progress.

Indeed, we think having improved the previous **Version N°2** with the following new elements which will be developed in the paragraphs of the study.

- **We consider now the weighted averages** (first moment) of the different parameters characterising the network (information flux, element transfer function,…). This introduces **probabilities** (refer statistic mechanics) and could, may be, allow a **checking** by experiments. (An interesting and surprising result is that, for a given network, the average of information flux is constant!)

- We give **an expression of the general State vector (wave function) of the network as a combination of 3 functions.**
- We have considered **2 expressions of the conactance** (the *Lagrangian* in Physics). Though the two are **invariant by SU(3)** Lie group of symmetry transformations and also by SU(2) in a representation of order 1, we cannot insure that these groups are valid; indeed, we think that, at the present stage, **only experiments** could lead to discover invariants (like strangeness in physics for ex.) which allow us to find the true symmetry groups. In addition to the computation of the Lagrange equations governing the evolution of the network, this could, in the future, be used to **classify the different types of networks**.
- We have defined 2 **new observables** (average of information flux and activity of the network). From our point of view, **these two observables could be measured** by experience and so this would give **the possibility of validate or invalidate the theory.**

To conclude this introduction, I would like to precise a point: I have a scientific formation (French engineer) and I have spent my professional career in industry. Now I am retired from many years ago. So I do not belong to any laboratory or research organism.

I am very interested by the behaviour of intelligent networks and particularly by the **brain**; and, as I took courses of very "just so" level about Relativity and Quantum mechanics, I had an idea: **Why not, by analogy, consider a network connected to its environment similar to an electron under a magnetic field influence for example and why not be inspired by the methods, the formalism of quantum mechanics?**

So I published my ideas (**Version N°2**) in view of exchange on these subjects with people or organisms interested. I am conscious that a pure theoretical study is not as mediatic as the impressive results obtained by projects like the *Human brain project* or *Artificial Intelligence*. But, may be, it will be necessary, later on, to elaborate a network general theory which will **explain scientifically** the results of all these studies…..

I am also conscious that, even if the idea was proved good, I have only touched lightly the topic; **Particularly, starting from the same hypothesis I did, other people could elaborate different quantum mathematical formalism**. So my best wish is that people more competent than me in *Quantum fields theory*, mathematics, neurobiology, computing, etc…and interested by the idea improve significantly the first outlines I have proposed.

If you are interested to exchange with me I would be very pleased to receive your comments.

My mail address is:

piniello.reseau@orange.fr

# CHAPTER I A BRIEF REMINDER OF VERSION N°2

**Note:** This reminder is a very brief summary of Version N°2 which has been issued in French language and comprises 46 pages.

## *I1 Definitions-Block diagrams- Notations.*

### I11 Network. A network is a set of elements

each of them being connected at least to another by an information relation
some of them having an information relation with the network environment
This set exchanges energy with this environment.

A network organizes itself ("**autoorganization**").

### I12 Definitions and Notations.

**Inlet elements**

- Elements which receive information from the environment.

Their number is $N_e$. (e as entry)

- Summation of inlet elements: $\sum_i$ with $i$=1 to $i$=$N_e$
- Each element can receive (n-N) data from the environment; they are called $f_1^i$ , $f_2^i$ ,.... $f_j^i$ ,... $f_{(n-N)}^i$

**Intermediate elements**

- Elements without any direct contact with the environment. Their number is $N_m$ ( m as middle)
- Summation of intermediate elements: $\sum_i$ with $i$=($N_e$+1) to $i$=($N_e$+$N_m$)

**Outlet elements**

- Elements which provide information to the environment. Their number is $N_s$ (s as "sortie" in French)

- At every outlet element is connected a *module* computing the difference between the information provided to the environment and the information that this element would have to provide (*set point*). Example : $l_N - s_N$
- Summation of outlet elements: $\sum_i$ with $i = N_e + N_m + 1$ to $i = N = N_e + N_m + N_s$
- The data provided by outlet elements must be equal to the *set points* imposed by environment.

**Each network element $E_i$**

- Receives *n* inputs
- Provides only one output $l_i$, which is a scalar (product of a line matrix by a column matrix)

The network includes N elements $E_i$ with $N = N_e + N_m + N_s$

**Free network**: Network which has no information connection with its environment.

**Connected network:** Network which has one or several information connections with its environment. These can

- Either be supplied by the environment and be received by the network (the $f_k^i$, input data)
- Or be supplied by the network and be received by the environment (output data)
- The environment can also set an output information to the network (the $s_i$ )

**Symbols**

$l_i$ **:** output information flow of the element $E_i$

$f_j^i$ : j[th] information flow provided by the environment and received by the inlet element $E_i$

$s_i$ : output information flow set by the environment (set point) to the outlet element $E_i$.

For this element, one want: $l_i = s_i$, and the quantities $(l_i - s_i)$ are re-injected as inputs to all the elements of the network

At the end of the apprenticeship period, we must obtain: $l_i = s_i$

$C$ : **Conactance** (The *Lagrangian* in QFT)

$\dot{t}_i^j = \frac{dt_i^j}{dt}$ : derivative of $t_i^j$ with respect to time

$\dot{l}_i = \frac{dl_i}{dt}$ : derivative of $l_i$ with respect to time

$\sum_{i=1}^{i=N}$ : summations ; we have not used the Einstein convention for the summations so as to no disturb the readers no familiar with this type of notation.

**Line Matrix symbolizing the *transfer function* of the element $E_i$ (**due to the fact that each element has only one output)

$$T_i = \left| a_i^1 \, a_i^2 \cdots a_i^{(n-N)} t_i^1 \cdots t_i^j \cdots t_i^N \right|$$

**Column Matrix symbolizing the inputs into element** $E_i$

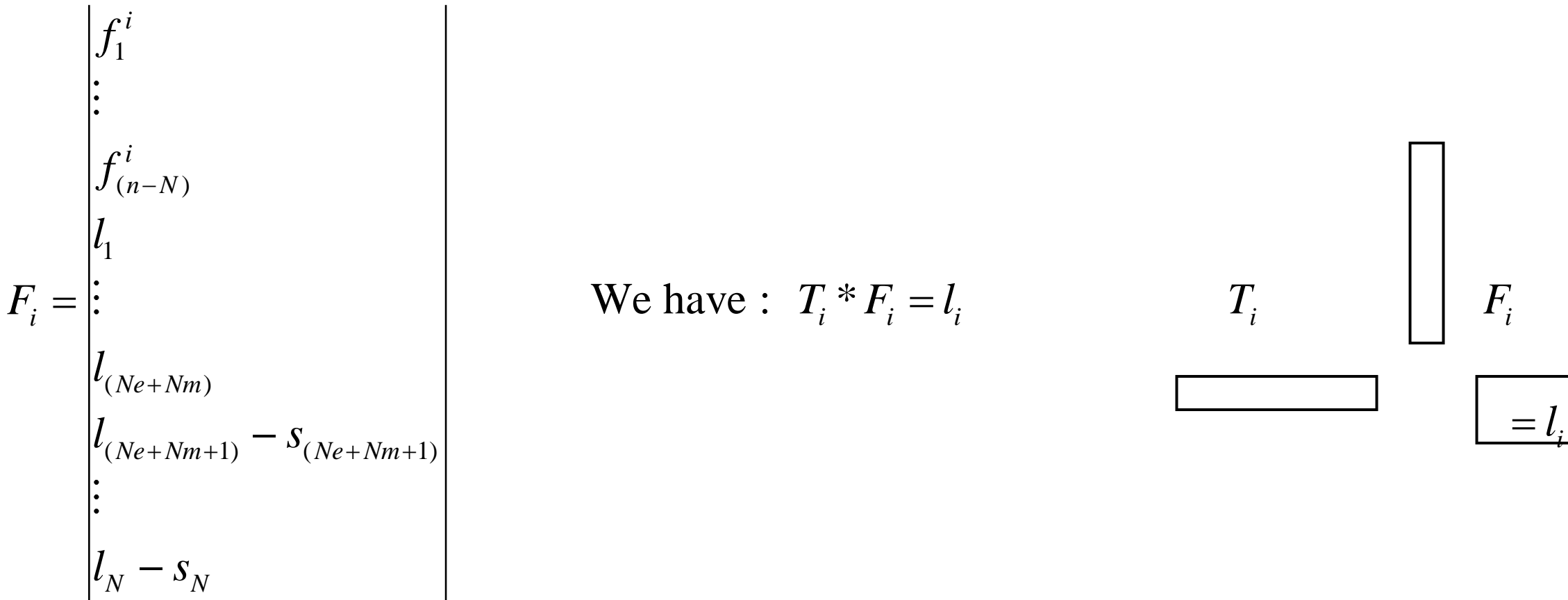

$n$ : dimension of $T_i$ and of $F_i$

**I13 Block diagrams. See Figures I1 and I2.**

Figure I1 Simplified block diagram of a network

Figure I2 Block diagram of an element $E_i$ of the network

(back to II1)

$s_1 - s_{10} = 0$

$s_2 - s_{20} = 0$

## Purpose of learning

$s_3 - s_{30} = 0$

when one imposes $e_1, e_2, e_3, e_4$

l’extérieur <u>à comparer</u>

**$S_2$**

avec les signaux que l’on

désire obtenir

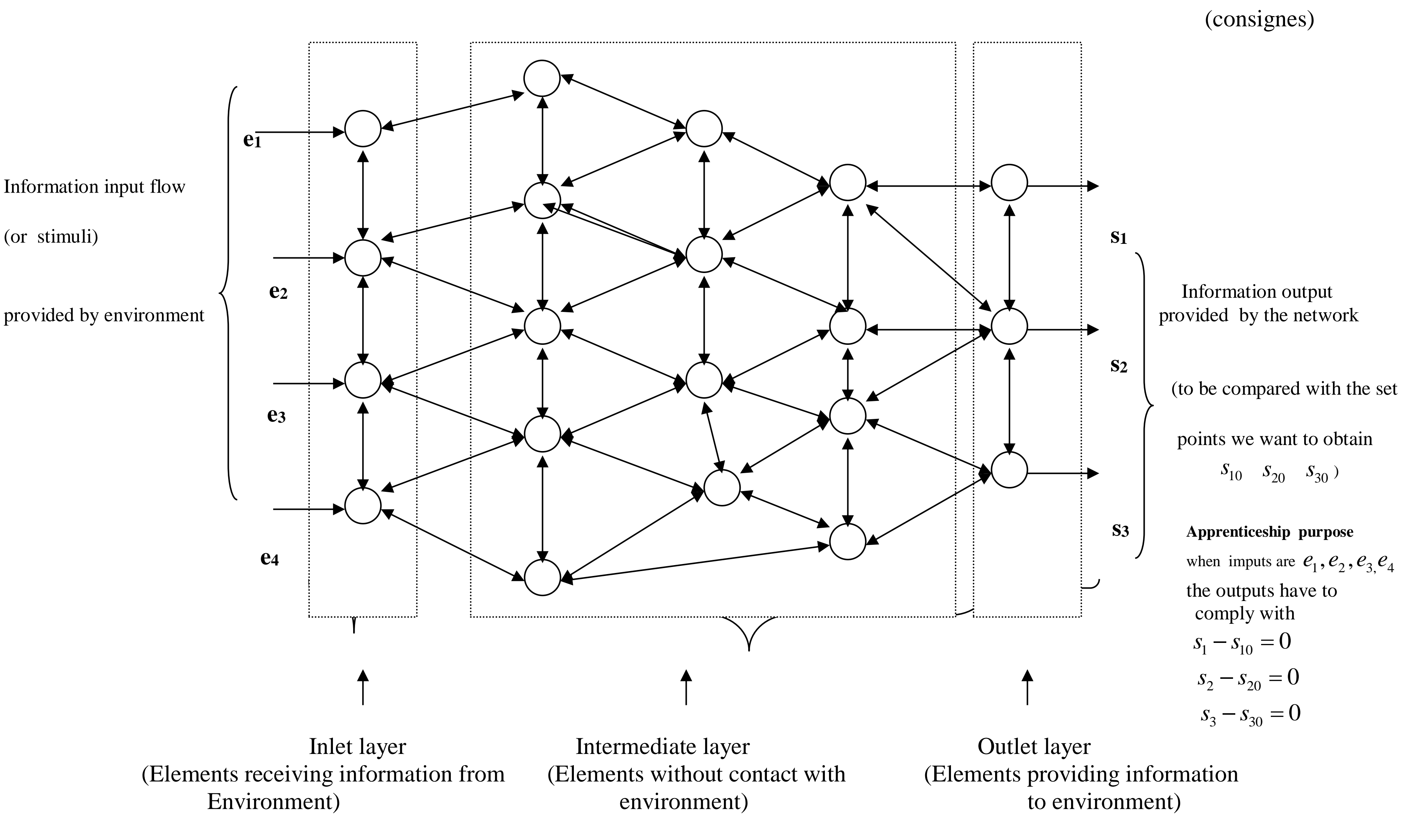

**Figure I1** **Simplified block diagram of a network** **Note:** All the connections between elements are not represented.

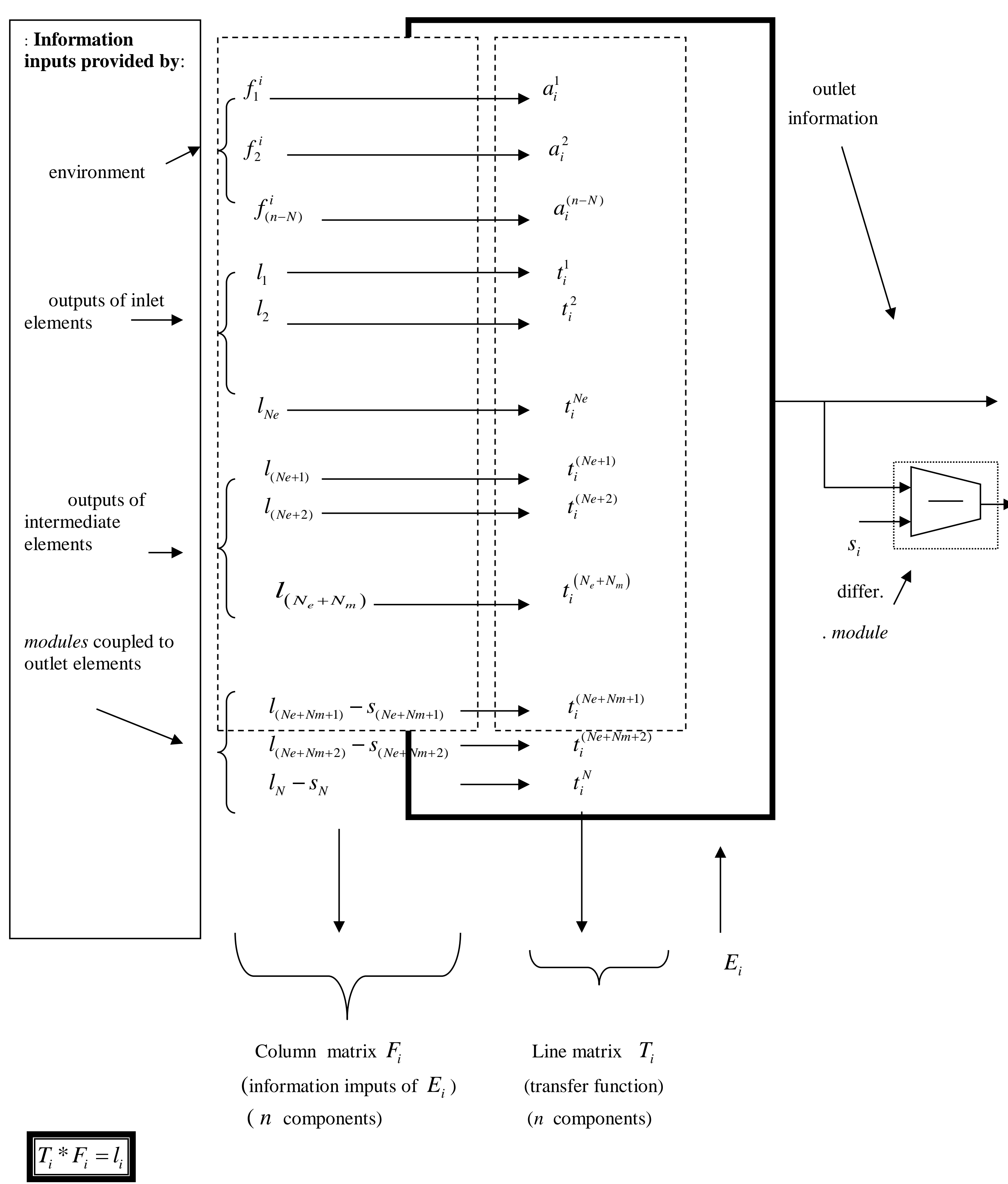

**Figure I-2** **Block diagram of an element $E_i$ of the network**

## I2 Principles-Formalism-Method.

- When the free network is connected with its environment, it can be considered as immersed in an information field, by analogy with a particle (an electron) submitted to a quantum field (electromagnetic field).
- So the formalism and the methods are inspired from those of *Quantum field theory* (Principle of stationary action, gauge fields, invariance by symmetry transformations…).
- Mainly we have to elaborate a ***conactance*** (the equivalent of the Lagrangian) which will be invariant under a symmetry transformation of the system, and deduce the Lagrange equations which give the evolution of the network.
- **Note**: Conactance is an invented “French” word; see [French Version 2, par. II 4] and Note [4] for more explanations. Conactance is expressed with information units while *Lagrangian* is expressed with energy units.

## I3 Conactance and Invariance.

- The ***conactance*** of the connected network is:

(I1)
$$C(t)=\int_0^t \sum_{i=1}^{i=N}\left[\sum_{k=1}^{k=n-N} f_k^i \dot{a}_i^{k2} + \sum_{k=1}^{k=N} l_k \dot{t}_i^{k2} - \sum_{k=(N_e+N_m+1)}^{k=N} s_k \dot{t}_i^{k2}\right]dt$$

and the ***stationary action*** is expressed by:

(I2)
$$S=\int_0^{t_1} C dt$$

- To keep invariant the *conactance* and the *flows balance* at each element, we have to consider a symmetry transformation which takes into account
  **simultaneously** a ***Permutation group*** transformation
  and a ***gauge change*** of the $f_k^i$ and the $s_i$ (the *information field* is considered as a *gauge field*)
- The connection of the free network to an information field can be interpreted on two ways:
  1) Physically, it connects the network to the external field which dictates the constraints during the apprenticeship period.
  2) Mathematically, it provides the invariance of the conactance under the symmetry transformation as seen above.

## I4 Flows balance at each element $E_i$ and computation of $l_i$.

- At any time the **balance of information flows** of each element $E_i$ drives to the following system of equations

**(I3)**

$$
\begin{array}{c}
\left(t_1^1-1\right)l_1+\cdots+t_1^i l_i+t_1^j l_j+t_1^k l_k+\cdots+t_1^N l_N=-f_1^1 a_1^1-\cdots-f_l^1 a_1^l-\cdots-f_{(n-N)}^1 a_1^{(n-N)}+t_1^p s_p+\cdots+t_1^N s_N \\
\cdots\cdots\cdots\cdots\cdots\cdots\cdots\cdots\cdots\cdots\cdots\cdots\cdots\cdots\cdots=\cdots\cdots\cdots\cdots\cdots\cdots\cdots\cdots\cdots\cdots\cdots\cdots\cdots\cdots\cdots\cdots\cdots\cdots\cdots\cdots \\
t_i^1 l_1+\cdots+\left(t_i^i-1\right)l_i+t_i^j l_j+t_i^k l_k+\cdots+t_i^N l_N=-f_1^i a_i^1-\cdots-f_l^i a_i^l-\cdots-f_{(n-N)}^i a_i^{(n-N)}+t_i^p s_p+\cdots+t_i^N s_N \\
t_j^1 l_1+\cdots+t_j^i l_i+\left(t_j^j-1\right)l_j+t_j^k l_k+\cdots+t_j^N l_N=-f_1^j a_j^1-\cdots-f_l^j a_j^l-\cdots-f_{(n-N)}^j a_j^{(n-N)}+t_j^p s_p+\cdots+t_j^N s_N \\
t_k^1 l_1+\cdots+t_k^i l_i+t_k^j l_j+\left(t_k^k-1\right)l_k+\cdots+t_k^N l_N=-f_1^k a_k^1-\cdots-f_l^k a_k^l-\cdots-f_{(n-N)}^k a_k^{(n-N)}+t_k^p s_p+\cdots+t_k^N s_N \\
\cdots\cdots\cdots\cdots\cdots\cdots\cdots\cdots\cdots\cdots\cdots\cdots\cdots\cdots\cdots=\cdots\cdots\cdots\cdots\cdots\cdots\cdots\cdots\cdots\cdots\cdots\cdots\cdots\cdots\cdots\cdots\cdots\cdots\cdots \\
t_N^1 l_1+\cdots+t_N^i l_i+t_N^j l_j+t_N^k l_k+\cdots+\left(t_N^N-1\right)l_N=-f_1^N a_N^1-\cdots-f_l^N a_N^l-\cdots-f_{(n-N)}^N a_N^{(n-N)}+t_N^p s_p+\cdots+t_N^N s_N
\end{array}
$$

Using this system we can compute each $l_i$

**(I4)**

$$
l_i=\frac{\Delta_i}{\Delta_0}=\frac{\begin{vmatrix}
\left(t_1^1-1\right) & \cdots-\left(f_1^1 a_1^1+\cdots+f_l^1 a_1^l+\cdots+f_q^1 a_1^q-t_1^p s_p-\cdots-t_1^N s_N\right) & t_1^j & t_1^k & \cdots t_1^N \\
\cdots & \cdots\cdots\cdots\cdots\cdots & \cdots & \cdots & \cdots \\
t_i^1 & \cdots-\left(f_1^i a_i^1+\cdots+f_l^i a_i^l+\cdots+f_q^i a_i^q-t_i^p s_p-\cdots-t_i^N s_N\right) & t_i^j & t_i^k & \cdots t_i^N \\
t_j^1 & \cdots-\left(f_1^j a_j^1+\cdots+f_l^j a_j^l+\cdots+f_q^j a_j^q-t_j^p s_p-\cdots-t_j^N s_N\right) & \left(t_j^j-1\right) & t_j^k & \cdots t_j^N \\
t_k^1 & \cdots-\left(f_1^k a_k^1+\cdots+f_l^k a_k^l+\cdots+f_q^k a_k^q-t_k^p s_p-\cdots-t_k^N s_N\right) & t_k^j & \left(t_k^k-1\right) & \cdots t_k^N \\
\cdots & \cdots\cdots\cdots\cdots\cdots & \cdots & \cdots & \cdots \\
t_N^1 & \cdots-\left(f_1^N a_N^1+\cdots+f_l^N a_N^l+\cdots+f_q^N a_N^q-t_N^p s_p-\cdots-t_N^N s_N\right) & t_N^j & t_N^k & \cdots\left(t_N^N-1\right)
\end{vmatrix}}{\begin{vmatrix}
\left(t_1^1-1\right) & \cdots t_1^i & t_1^j & t_1^k & \cdots t_1^N \\
\cdots\cdots & \cdots\cdots & \cdots\cdots & \cdots\cdots & \cdots\cdots \\
t_i^1 & \cdots\left(t_i^i-1\right) & t_i^j & t_i^k & \cdots t_i^N \\
t_j^1 & \cdots t_j^i & \left(t_j^j-1\right) & t_j^k & \cdots t_j^N \\
t_k^1 & \cdots t_k^i & t_k^j & \left(t_k^k-1\right) & \cdots t_k^N \\
\cdots & \cdots\cdots & \cdots\cdots & \cdots\cdots & \cdots\cdots \\
t_N^1 & \cdots t_N^i & t_N^j & t_N^k & \cdots\left(t_N^N-1\right)
\end{vmatrix}}
$$

## *I5 Lagrange equations.*

We have to determine the functions $a_i^k(t), t_i^k(t), l_k(t)$ taking into account that each of them , if it is called $f(t)$, has to comply with the following Lagrange equation

**(I5)**
$$\frac{\partial C}{\partial f} - \frac{d}{dt}\frac{\partial C}{\partial \dot{f}} = 0 \quad \textbf{with} \quad \dot{f} = \frac{df}{dt}$$

The results are: ( for more details, see [French Version 2, par. IV 8])

**Functions** $a_i^k$ : They are constants.

**Functions** $t_i^k$ : They are also constants for the values of $i = 1$ to $N$ and $k = (N_e + N_m + 1)$ to $N$ .
On a general manner the transfer function components which receive information flows dictated by environment are **constants** whose value is that they had at the time when the free network was connected to this environment.

**Other functions**

$t_i^k$ **:** They must satisfy the differential equation

**(I6)**
$$\left[ 2l_k \ddot{t}_i^k + \dot{l}_k \dot{t}_i^k \right] = 0 \qquad \textbf{or}$$

**(I7)**
$$\boxed{\Delta_k \dot{t}_i^{k\,2} = \Delta_0 K_{ik}}$$

$K_{ik}$ being a constant imposed by limits conditions

$\Delta_0$ and $\Delta_k$ being the determinants of expression **(I4)** which gives the

functions $l_k$ by $l_k = \dfrac{\Delta_k}{\Delta_0}$

## *I6 Suggested solutions.*

It is obvious that, due to the number of variables, all these differential equations are very difficult to solve.

Nevertheless, considering that the solutions have to take into account two characteristics (memory and weakening), we suggest two shapes.

a) **For memory** (when the network receives the inputs used during its apprenticeship, it must furnish the outputs imposed )

**(I8)**
$$l_k(t) = s_k e^{-\sum \left| f_k^i - {}_a f_k^i \right|} v(t)$$

if the $l_k$ can be described by $l_k = qv(t)$

b) **For weakening** (when a connection is not often used its influence must be reduced, by means of the corresponding $t_i^k$ **).**

**(I9)** $$t_i^k(t) = \frac{T_{ik}}{t}\left[\int_0^t l_k dt\right] u(t)$$

if the $t_i^k$ can be described by $t_i^k = pu(t)$ and $\langle l_k \rangle_0^t = \frac{1}{t}\int_0^t l_k dt$

(For more details, see French Version 2, par. IV 9)

## *I7 Suggested research ways-Possible improvements.*

It could be interesting to search

Other symmetry transformations
Other expression of the conactance
Other *observables* more accessible than the $t_i^k$ and the $l_k$
A representative State vector (Wavefunction)
May be *invariants*, in view of a classification
An equivalent of the Shrödinger equation in *Quantum mechanics*
and so on…

Furthermore, **the most important problem to solve is the connection of several networks together, each of them having been subject to learning: we consider this is a fundamental point to understand how the brain works**; here we suggest methods used in *Quantum mechanics* (for instance: two particles interaction).

But the major inconvenient of Version 2 is the too great number of variables (the $l_k$ and the $t_i^k$ **)** and consequently the too great number of differential equations**,** which prevent us to have a "macroscopic" view of the network evolution. Also, considering this considerable number, a **statistic approach** seems to be judicious.

The purpose of the following Version 5 is to improve these insufficiencies of the Version 2

# CHAPTER II. VERSION 5 OF THE THEORY

## II1 Definitions-Notations- Block diagrams.

They are identical to those of Version 2 (see [par. **I1**])

## II2 Principles- Method- Formalism.

- As in Version 2, a free network connected to its environment can be considered as immerged in an information field, by analogy with a particle submitted to an electromagnetic field.
- We will use also the formalism of the QFT (*quantum field theory*), and particularly that of gauge fields, principle of stationary action, invariance under symmetry transformations…
- But we try to improve by adding new concepts as
  - a *State vector* (or *wave function*) for the system, combination of 3 functions
  - two possible expressions of the *conactance*
  - new *observables*, which could be measured by experiments
  - and, mainly, the replacement of the variables $t_i^k$ and $l_k$ by their **weighted averages;** this introduces **probabilities.**

## II3 Weighted averages of $t_i^k$ and $l_k$.

- We consider that, **at a given time** $t$**,** the $t_i^k$ can be considered as a **set of random variables.**

Let us call

$$T = \prec t_i^k \succ = \sum p_i^k t_i^k$$ the weighted average (1st moment) of the $t_i^k$

$p_i^k$ being the probability that, **at a given time** $t$**,** the variable $t_i^k$ takes the value $t_i^k$ **.**

$T$ is a function of time $t$

In the same way, we can write

$$L = \prec l_k \succ = \sum p_k l_k$$ the weighted average (1st moment) of the $l_k$

$p_k$ being the probability that, **at a given time** $t$**,** the variable $l_k$ takes the value $l_k$ **.**

$L$ is a function of time $t$

- On this basis, we can say that

$$\dot{T} = \prec \dot{t}_i^k \succ = \sum p_{ik} \dot{t}_i^k$$ because the probability to have $\dot{t}_i^k$ is the same than that to have $t_i^k$ **.**

But $\prec \dot{t}_i^{k2} \succ$ is not equal to $\dot{T}^2$ . However, as we want simplify and keep expressions yet used in Version 2, like $l_k \dot{t}_i^{k2}$ , we will use the term $L\dot{T}^2$ in lieu of $L \prec \dot{t}_i^{k2} \succ$. In the future and if necessary, it will be possible to consider the true value of $\prec \dot{t}_i^{k2} \succ$

- **Important: Value of $L$.**

The value of any $l_k$ is given by the quotient of determinants $\frac{\Delta_k}{\Delta_0}$ ((see [ equation **(I4)**]); this is determined by the Cramer system indicated in **(I3).**
The probability to get this $l_k$ is the same than to get the determinants $\Delta_k$ and $\Delta_0$, which, finally, is the probability to obtain, at a given time $t$, the system of equations **(I3).**

Therefore, as the probability to obtain any $l_k$ is the same than the probability to obtain the system **(I3), the probabilities of all the $l_k$ are equal,** the common value being $\frac{1}{N}$.

So we can write **(II1)** $L = \prec l_k \succ = \frac{1}{N}\sum l_k = \frac{1}{N}\sum \frac{\Delta_k}{\Delta_0} = \frac{1}{N\Delta_0}\sum \Delta_k$

**But** $\sum \Delta_k = (N-1)\Delta_0$ **therefore (II2)** $\boxed{L = \frac{(N-1)\Delta_0}{N\Delta_0} = \frac{(N-1)}{N}}$

**Thus, for a given network comprising $N$ elements, the weighted average of the information produced by the output of each of these elements is a constant which does not depend on the time $t$.**

| This is an important result because we can consider $L$ as a constant, may be as an invariant in a possible classification of the networks, and also, mainly, as a value which could be tested by experiments. |
|---|

## *II4 State vector (Wavefunction) of the system.*

- **Preliminary: Symbols adopted.**

$F$ means weighted average of the $f_k^i$ of inlet elements $i=1$ to $N_e$ and $k=1$ to $(n-N)$

$A_e$ means weighted average of the $a_i^k$ of inlet elements $i=1$ to $N_e$ and $k=1$ to $(n-N)$

$L$ means weighted average of the $l_i$ $i=1$ to $N$

$T_e$ means weighted average of the $t_i^j$ of inlet elements $i=1$ to $N_e$ and $j=1$ to $N$

$T_m$ means weighted average of the $t_i^j$ of middle elements $i=(N_e+1)$ to $(N_e+N_m)$
and $j=1$ to $N$

$T_s$ means weighted average of the $t_i^j$ of outlet elements $i=(N_e+N_m+1)$ to $N$
and $j=1$ to $N$

$T_{se}$ means weighted average of the $t_i^j$ of inlet elements $i=1$ to $N_e$
and $j=(N_e+N_m+1)$ to $N$

$T_{sm}$ means weighted average of the $t_i^j$ of middle elements $i=(N_e+1)$ to $(N_e+N_m)$
and $j=(N_e+N_m+1)$ to $N$

$T_{ss}$ means weighted average of the $t_i^j$ of outlet elements $i = (N_e + N_m + 1)$ to $N$

and $j = (N_e + N_m + 1)$ to $N$

$S$ means weighted average of the $s_i$ $i = (N_e + N_m + 1)$ to $N$

- **State vector (Wavefunction)**

We propose a state vector $\Phi$ of the system, constituted by 3 functions ; each of them corresponds respectively with one of the parts of the network (inlet, middle, outlet elements).
Though this expression of the state vector be purely intuitive , we will see further that the 3 functions corresponds to an *observable*. Likewise, this expression of Φ will be probably more adapted to the network after its learning (see or return to **[II9 2]**).

**(II3)**

$$\Phi = \begin{vmatrix} \varphi_e \\ \varphi_m \\ \varphi_s \end{vmatrix} = \begin{vmatrix} \frac{N_e}{N}\left( F\dot{A}_e^{\,2} + L\dot{T}_e^{\,2} - S\dot{T}_{se}^{\,2} \right) \\ \frac{N_m}{N}\left( L\dot{T}_m^{\,2} - S\dot{T}_{sm}^{\,2} \right) \\ \frac{N_s}{N}\left( L\dot{T}_s^{\,2} - S\dot{T}_{ss}^{\,2} \right) \end{vmatrix}$$

## *II5 Conactance (Lagrangian in* Quantum mechanics*)*

- We will consider 2 ways for the conactance expression:
  1) One which takes again that used in Version 2, but with weighted averages.
  2) Another inspired by a general expression used in Quantum mechanics (kinetic term plus potential term).

$1^{st}$ case. **(II5)** (By analogy with Version 2)

$$C = \int_0^t (\varphi_e + \varphi_m + \varphi_s) dt = \int_0^t \left[ \frac{N_e}{N}\left( F\dot{A}_e^{\,2} + L\dot{T}_e^{\,2} - S\dot{T}_{se}^{\,2} \right) + \frac{N_m}{N}\left( L\dot{T}_m^{\,2} - S\dot{T}_{sm}^{\,2} \right) + \frac{N_s}{N}\left( L\dot{T}_s^{\,2} - S\dot{T}_{ss}^{\,2} \right) \right] dt$$

$2^{nd}$case. **(II6)**

$$C = \partial_t \Phi^* \partial_t \Phi - k\Phi^*\Phi \quad = \dot{\Phi}^* \dot{\Phi} - k\Phi^*\Phi$$

$\Phi^*$ **being the transposed matrix of** $\Phi$

**As (II7)** $$\partial_t \Phi = \dot{\Phi} = \begin{vmatrix} \partial_t \varphi_e \\ \partial_t \varphi_m \\ \partial_t \varphi s \end{vmatrix} = \begin{vmatrix} \dot{\varphi}_e \\ \dot{\varphi}_m \\ \dot{\varphi}_s \end{vmatrix} = \begin{vmatrix} \frac{N_e}{N}\left(2F\,\dot{A}_e\,\ddot{A}_e + 2L\dot{T}_e\,\ddot{T}_e - 2S\,\dot{T}_{se}\,\ddot{T}_{se}\right) \\ \frac{N_m}{N}\left(2L\dot{T}_m\,\ddot{T}_m - 2S\,\dot{T}_{sm}\,\ddot{T}_{sm}\right) \\ \frac{N_s}{N}\left(2L\dot{T}_s\,\ddot{T}_s - 2S\,\dot{T}_{ss}\,\ddot{T}_{ss}\right) \end{vmatrix}$$

**and (II8)** $$\Phi^*\Phi = \varphi_e^{\,2} + \varphi_m^{\,2} + \varphi_s^{\,2}$$

**and (II9)** $$\dot{\Phi}^*\,\dot{\Phi} = \dot{\varphi}_e^{\,2} + \dot{\varphi}_m^{\,2} + \dot{\varphi}_s^{\,2}$$

**after computation we obtain (II10)**

$$
\begin{aligned}
C = {} & 4\frac{N_e^{\,2}}{N^2}\left(F^2\,\dot{A}_e^{\,2}\,\ddot{A}_e^{\,2} + L^2\,\dot{T}_e^{\,2}\,\ddot{T}_e^{\,2} + S^2\,\dot{T}_{se}^{\,2}\,\ddot{T}_{se}^{\,2} + 2FL\,\dot{A}_e\,\ddot{A}_e\,\dot{T}_e\,\ddot{T}_e - 2FS\,\dot{A}_e\,\ddot{A}_e\,\dot{T}_{se}\,\ddot{T}_{se} - 2LS\,\dot{T}_e\,\ddot{T}_e\,\dot{T}_{se}\,\ddot{T}_{se}\right) \\
& +4\frac{N_m^{\,2}}{N^2}\left(L^2\,\dot{T}_m^{\,2}\,\ddot{T}_m^{\,2} + S^2\,\dot{T}_{sm}^{\,2}\,\ddot{T}_{sm}^{\,2} - 2LS\,\dot{T}_m\,\ddot{T}_m\,\dot{T}_{sm}\,\ddot{T}_{sm}\right) \\
& +4\frac{N_s^{\,2}}{N^2}\left(L^2\,\dot{T}_s^{\,2}\,\ddot{T}_s^{\,2} + S^2\,\dot{T}_{ss}^{\,2}\,\ddot{T}_{ss}^{\,2} - 2LS\,\dot{T}_s\,\ddot{T}_s\,\dot{T}_{ss}\,\ddot{T}_{ss}\right) \\
& -k\frac{N_e^{\,2}}{N^2}\left(F^2\,\dot{A}_e^{\,4} + L^2\,\dot{T}_e^{\,4} + S^2\,\dot{T}_{se}^{\,4} + 2FL\,\dot{A}_e^{\,2}\,\dot{T}_e^{\,2} - 2FS\,\dot{A}_e^{\,2}\,\dot{T}_{se}^{\,2} - 2LS\,\dot{T}_e^{\,2}\,\dot{T}_{se}^{\,2}\right) \\
& -k\frac{N_m^{\,2}}{N^2}\left(L^2\,\dot{T}_m^{\,4} + S^2\,\dot{T}_{sm}^{\,4} - 2LS\,\dot{T}_m^{\,2}\,\dot{T}_{sm}^{\,2}\right) \\
& -k\frac{N_s^{\,2}}{N^2}\left(L^2\,\dot{T}_s^{\,4} + S^2\,\dot{T}_{ss}^{\,4} - 2LS\,\dot{T}_s^{\,2}\,\dot{T}_{ss}^{\,2}\right)
\end{aligned}
$$

## *II6 Invariance- Symmetry transformations group.*

- Each system is characterised by one or several symmetry transformations; using them, we can

  a) Classify the system (static point of view)

  b) Explain the interactions with other systems (dynamic point of view)

- For example, in *Quantum mechanics*, some particles, such as baryons, have been classified inside general families of particles (multiplets); this was possible using symmetry groups SU(2) and SU(3), related to *invariants* like *strangeness* or *hypercharge*.

In the same way the interactions between particles were established by writing a Lagrangian invariant under these symmetry transformations.

- However it is important to not forget that all this construction has been possible because many experiments were performed about particles, and the properties of *strangeness* or *hypercharge*, for instance, were discovered little by little. **It was only after these experimental discoveries that the theory, using SU(2) and SU(3), could be finalised**.

- But in our case, so far we know, equivalent experiments for networks had not been implemented; so it is very difficult, for the present study which is purely theoretical, to propose symmetry groups capable of keeping the *conactance* invariant and classifying networks by families.

We hope that the numerous studies about brain or Artificial Intelligence carried out currently in many laboratories will discover new properties (*invariants, possible multiplets*, etc…); then, if the hypothesis of this study are valuable, our theory could be completed.

- Nevertheless, we can say that

a) The **Permutation group**, used in Version 2, remains a symmetric group for the network

b) The SU(3) group, applied to the 2 types of *conactance,* keep them invariant (and we think also SU(2) in a representation of order 1)

c) Mainly, **the weighted average** of the information produced by the N elements of the network, $L$ , is **an invariant** (see **(II2)** $L=\frac{N-1}{N}$). This property is very interesting, because probably it could be measured by experiments.

## II7 Lagrange equations.

- We have to determine the functions $A_e$, $T_{\bullet}$, $T_{\bullet\bullet}$, as named at the beginning of par. **II3,** taking into account that each of them , if it is called $f(t)$, has to comply with the following Lagrange equation

**(II11)** $$\frac{\partial C}{\partial f}-\frac{d}{dt}\frac{\partial C}{\partial \dot{f}}=0 \quad \textbf{with} \quad \dot{f}=\frac{df}{dt}$$

- 1st case The conactance is given by expression **[ (II5)]**

We have only derivatives, so the terms $\frac{\partial C}{\partial f}$ have not to be considered; the terms to be taken into account are:

**(II12)** $$\frac{d}{dt}\bullet\frac{\partial C}{\partial \dot{f}}=0$$

and more precisely their form is:

**(II13)** $$\frac{d}{dt}\bullet k\frac{\partial\int_0^t f dt}{\partial \dot{f}}=0$$

For instance, $T_e$ will be given by:

**(II14)** $$\frac{d}{dt} \bullet \frac{N_e}{N} \int_0^t L \dot{T}_e^{\,2}\, dt = 0$$

To express this equation under a differential equation form we use a general computation carried out in the **[Note [6]] "Expression of Lagrange equations" of our French Version 2**.

**Given an equation (II15)** $$\frac{d}{dt} \bullet \frac{\partial \int_0^t p(t) \left[ \dot{q}(t) \right]^2}{\partial \dot{q}(t)} = 0$$ **where** $p(t)$ **and** $q(t)$ **are functions of** $t$

we find that this expression can be written

**(II16)** $$\frac{\dot{q}(t)}{\ddot{q}(t)} \left[ 2 p(t) \ddot{q}(t) + \dot{q}(t) \dot{p}(t) \right] = 0$$

If $p$ is a constant, like in **(III13)** or **(III14),** $\dot{p}(t) = 0$ and (III16) becomes

**(II17)** $$2 p \dot{q}(t) = 0$$

So the general equation **(III13)** can be written

**(II18)** $$2k \dot{f} = 0$$

That means that $f$ is a constant.

<u>Note</u>: We can also compute $\int_0^t \dot{T}_e^{\,2}\, dt$ by integrating by parts; we have

**(II19)** $$\int_0^t \dot{T}_e^{\,2}\, dt = \left| T_e \dot{T}_e \right|_0^t - \int_0^t T_e \ddot{T}_e\, dt$$

and **(II14)** becomes $T_e \ddot{T}_e + \dot{T}_e^{\,2} - T_e \ddot{T}_e = \dot{T}_e^{\,2} = 0$

Then $\dot{T}_e = 0$ and $T_e$ is a constant.

**That means that the weighted averages** $A_e$, $T_e$, $T_m$, $T_s$, $T_{se}$, $T_{sm}$, $T_{ss}$ **are constants.**

This is a very surprising result!

<u>Remarks:</u> 1) This result does not come from equal probabilities like for $L$

2) Its validity could be checked by experiments (cf paragraph on *observables*)

- <u>2nd case</u> The conactance is given by expression [ (II10)]

Here also, we have only derivatives, so the terms $\frac{\partial C}{\partial f}$ have not to be considered; the terms to be taken into account are:

$$\frac{d}{dt} \bullet \frac{\partial C}{\partial \dot{f}} = 0$$

**(II20)**

$$\frac{d}{dt} \bullet \frac{\partial C}{\partial \dot{A}_e} = 4 \left( \frac{N_e}{N} \right)^2 \left[ \begin{array}{l} 2F^2 \ddot{A}_e^{\,3} + 4F^2 \dot{A}_e \ddot{A}_e \dddot{A}_e + 2FL \dddot{A}_e \dot{T}_e \ddot{T}_e + 2FL \ddot{A}_e \ddot{T}_e^{\,2} + 2FL \ddot{A}_e \dot{T}_e \dddot{T}_e - 2FS \dddot{A}_e \dot{T}_{se} \ddot{T}_{se} - 2FS \ddot{A}_e \ddot{T}_{se}^{\,2} - 2FS \ddot{A}_e \dot{T}_{se} \dddot{T}_{se} \\ -k \left( 3F^2 \dot{A}_e^{\,2} \ddot{A}_e + FL \ddot{A}_e \dot{T}_e^{\,2} + 2FL \dot{A}_e \dot{T}_e \ddot{T}_e - FS \ddot{A}_e \dot{T}_{se}^{\,2} - 2FS \dot{A}_e \dot{T}_{se} \ddot{T}_{se} \right) \end{array} \right] = 0$$

**(II21)**

$$\frac{d}{dt}\bullet\frac{\partial C}{\partial \dot{T}_e}=4\left(\frac{N_e}{N}\right)^2\left[\begin{array}{l}2L^2\ddot{T}_e^{\,3}+4L^2\dot{T}_e\ddot{T}_e\dddot{T}_e+2FL\ddot{A}_e^{\,2}\ddot{T}_e+2FL\dot{A}_e\dddot{A}_e\ddot{T}_e+2FL\dot{A}_e\ddot{A}_e\dddot{T}_e-2LS\dddot{T}_e\dot{T}_{se}\ddot{T}_{se}-2LS\ddot{T}_e\ddot{T}_{se}^{\,2}-2LS\ddot{T}_e\dot{T}_{se}\dddot{T}_{se}\\ -k\left(3L^2\dot{T}_e^{\,2}\ddot{T}_e+2FL\dot{A}_e\ddot{A}_e\dot{T}_e+FL\dot{A}_e^{\,2}\ddot{T}_e-LS\ddot{T}_e\dot{T}_{se}^{\,2}-2LS\dot{T}_e\dot{T}_{se}\ddot{T}_{se}\right)\end{array}\right]=0$$

**(II22)**

$$\frac{d}{dt}\bullet\frac{\partial}{\partial \dot{T}_{se}}=4\left(\frac{N_e}{N}\right)^2\left[\begin{array}{l}2S^2\ddot{T}_{se}^{\,3}+4S^2\dot{T}_{se}\ddot{T}_{se}\dddot{T}_{se}-2FS\ddot{A}_e^{\,2}\ddot{T}_{se}-2FS\dot{A}_e\dddot{A}_e\ddot{T}_{se}-2FS\dot{A}_e\ddot{A}_e\dddot{T}_{se}-2LS\ddot{T}_e^{\,2}\ddot{T}_{se}-2LS\dot{T}_e\dddot{T}_e\ddot{T}_{se}-2LS\dot{T}_e\ddot{T}_e\dddot{T}_{se}\\ -k\left(3S^2\dot{T}_{se}^{\,2}\ddot{T}_{se}-2FS\dot{A}_e\ddot{A}_e\dot{T}_{se}-FS\dot{A}_e^{\,2}\ddot{T}_{se}-2LS\dot{T}_e\ddot{T}_e\dot{T}_{se}-LS\dot{T}_e^{\,2}\ddot{T}_{se}\right)\end{array}\right]=0$$

**(II23)**

$$\frac{d}{dt}\bullet\frac{\partial}{\partial \dot{T}_m}=4\left(\frac{N_m}{N}\right)^2\left[\begin{array}{l}2L^2\ddot{T}_m^{\,3}+4L^2\dot{T}_m\ddot{T}_m\dddot{T}_m-2LS\dddot{T}_m\dot{T}_{sm}\ddot{T}_{sm}-2LS\ddot{T}_m\ddot{T}_{sm}^{\,2}-2LS\ddot{T}_m\dot{T}_{sm}\ddot{T}_{sm}\\ -k\left(3L^2\dot{T}_m^{\,2}\ddot{T}_m-LS\ddot{T}_m\dot{T}_{sm}^{\,2}-2LS\dot{T}_m\dot{T}_{sm}\ddot{T}_{sm}\right)\end{array}\right]=0$$

**(II24)**

$$\frac{d}{dt}\bullet\frac{\partial}{\dot{T}_{sm}}=4\left(\frac{N_m}{N}\right)^2\left[\begin{array}{l}2S^2\ddot{T}_{sm}^{\,3}+4S^2\dot{T}_{sm}\ddot{T}_{sm}\dddot{T}_{sm}-2LS\ddot{T}_m^{\,2}\ddot{T}_{sm}-2LS\dot{T}_m\dddot{T}_m\ddot{T}_{sm}-2LS\dot{T}_m\ddot{T}_m\dddot{T}_{sm}\\ -k\left(3S^2\dot{T}_{sm}^{\,2}\ddot{T}_{sm}-2LS\dot{T}_m\ddot{T}_m\dot{T}_{sm}-LS\dot{T}_m^{\,2}\ddot{T}_{sm}\right)\end{array}\right]=0$$

**(II25)**

$$\frac{d}{dt}\bullet\frac{\partial}{\partial \dot{T}_s}=4\left(\frac{N_s}{N}\right)^2\left[\begin{array}{l}2L^2\ddot{T}_s^{\,3}+4L^2\dot{T}_s\ddot{T}_s\dddot{T}_s-2LS\dddot{T}_s\dot{T}_{ss}\ddot{T}_{ss}-2LS\ddot{T}_s\ddot{T}_{ss}^{\,2}-2LS\ddot{T}_s\dot{T}_{ss}\dddot{T}_{ss}\\ -k\left(3L^2\dot{T}_s^{\,2}\ddot{T}_s-LS\ddot{T}_s\dot{T}_{ss}^{\,2}-2LS\dot{T}_s\dot{T}_{ss}\ddot{T}_{ss}\right)\end{array}\right]=0$$

**(II26)**

$$\frac{d}{dt}\bullet\frac{\partial}{\partial \dot{T}_{ss}}=4\left(\frac{N_s}{N}\right)^2\left[\begin{array}{l}2S^2\ddot{T}_{ss}^{\,3}+4S^2\dot{T}_{ss}\ddot{T}_{ss}\dddot{T}_{ss}-2LS\ddot{T}_s^{\,2}\ddot{T}_{ss}-2LS\dot{T}_s\dddot{T}_s\ddot{T}_{ss}-2LS\dot{T}_s\ddot{T}_s\dddot{T}_{ss}\\ -k\left(3S^2\dot{T}_{ss}^{\,2}\ddot{T}_{ss}-2LS\dot{T}_s\ddot{T}_s\dot{T}_{ss}-LS\dot{T}_s^{\,2}\ddot{T}_{ss}\right)\end{array}\right]=0$$

- **About solutions.**

We must solve a system of 7 non linear differential equations and find 7 functions; this seems difficult.

.

However we can note that if we apply to **(II20)** the transformations

$\dot{A}_e \rightarrow \dot{T}_{se}, \dot{T}_{se} \rightarrow \dot{T}_e, \dot{T}_e \rightarrow \dot{A}_e, F \rightarrow S, S \rightarrow L, L \rightarrow -F$ we meet again **(II22)**.

In the same way, if we apply to **(II26)** the transformations $\dot{T}_s \rightarrow \dot{T}_{ss}, \dot{T}_{ss} \rightarrow \dot{T}_s, S \rightarrow L, L \rightarrow S$ we meet again **(II25)**. Idem for **(II23)** and **(II24)**.

All this suggests that these differential equations have common forms and that the method for solving them would be identical. This is not surprising due to the fact that the 7 functions we want to find represent the evolution of elements of same nature; merely **these functions will differ by initial and final conditions and also by the presence of *L* or *F* or *S*.**

Another question is to know if one equation has to be solved by equalizing to zero each of its terms; (example: $4L^2 \dot{T}_s \ddot{T}_s \dddot{T}_s = 0$).
But if we are not able to solve this system, we are convinced that good mathematicians could do it….
Remark: Due to the great number of elements, we could consider continuous probability density functions, in lieu of discrete probabilities.

## *II8 Observables- Comparison with experiments.*

- **Any theory has to be confronted with experimental results.** That is to say that some characteristics of a studied system have to be measured (they are called ***observables***) so as to check if they coincide with the predictive results given by the theory.

- In our case we propose 2 observables:

  1) $L$ : ***the weighted average*** of the $l_i$ issued from the different $E_i$ .
  2) $A$ : ***the activity of the network*** with

$$A = \frac{N_e}{N}\left(F\dot{A}_e^{\,2} + L\dot{T}_e^{\,2} - S\dot{T}_{se}^{\,2}\right) + \frac{N_m}{N}\left(L\dot{T}_m^{\,2} - S\dot{T}_{sm}^{\,2}\right) + \frac{N_s}{N}\left(L\dot{T}_s^{\,2} - S\dot{T}_{ss}^{\,2}\right)$$

Note: May be we could omit the terms $\frac{N_i}{N}$.

- Why these 2 ***observables***?

1) For $L$ we have deduced that this quantity could be an **invariant** (see [par. **II3**] and [equation **(II 2)**]); it seems interesting to test it.
2) For $A$ we know that the activity of the brain can be displayed qualitatively; no doubt that in the future this characteristic will be perceived **quantitatively (**mainly by coloring**).**
3) **But above all we think that, due to the considerable number of studies currently carried out about the brain, the most sophisticated of networks, and artificial intelligence, it will be possible to measure many characteristics using new technologies (**such as, for example, quantity of information furnished by a neuron at its output).

## *II9 Future research ways.*

We think that we have to distinguish two domains:
**1) The first, when the network is in its learning period: it is the purpose of this study (Version 2 improved by Version 5). In spite of improvements brought by Version 5, mainly by considering weighted averages (which introduces probabilities) and by proposing two observables (Average information *L* and Activity *A*), we are conscious that it remains much work to progress.**

We certainly could progress by checking for example the dimensional homogeneity of equations, or by proposing other expressions for the *conactance* or the *state vector*, but we think **that major improvements will come from the considerable experimental researches carried out currently in many laboratories about the networks, and mainly the brain and artificial intelligence**. **Indeed, only experiments could allow us to find physical properties and particularly invariants which, taken up by theory, would precise which symmetry groups are capable of establish a classification of networks and also their interconnections.**

**2)The second, every bit as important as the first, when the network, after its learning period, is connected with other networks; this is the case for the brain which is a set of a considerable number of sub-networks connected together. Each of these sub-network has been subject to learning and is devoted to assume a biological function; as far as the brain is concerned, we remind that the interconnections of this great number of sub-networks generate the conscious mental states.**

**The problem is to know why and how these sub-networks connect together**.

By analogy with Quantum mechanics, we could try to use the methods giving the probabilities of interconnection of two particles (mainly Feynman diagrams; see [par. **VI3** "*Autres voies de recherche concernant le couplage de plusieurs réseaux*"] of the French Version 2); but this needs to know, particularly, their state vector and symmetry groups. About this subject, may be that the expression of space vector given in [paragraph **II4]** could be pertinent for it takes into account the memory of apprenticeship (terms *F*, *S*)….

# CONCLUSION

This **Version 5** which deals with the study of networks during their learning period, brings about, we think, **relevant improvements** with regard to Version 2: ***weighted averages*** (which introduce *probabilities*), ***observables*** (*Activity **A*** of the network and *Information flux average L*, which could **be compared with the results of experiments**), ***invariance*** of the observable *L*.

But we are convinced that to go further, it is necessary, as it has been done in Quantum mechanics (**experimental** discovery of invariants like *hypercharge* and *strangeness* for ex.), that **experiments** about networks, and particularly **the brain or those used in artificial intelligence**, be carried out, so as to discover characteristic **invariants**. Indeed such discoveries could lead to improve significantly the theory by introducing symmetry groups and vector state; so we could hope to establish a classification of networks and understand their interactions.

But we have only touched lightly on the problem of **interactions** (**connecting**) between several networks (or sub-networks) after they have succeeded their learning period.
Indeed, to speak about the **brain**, which is the most sophisticated network and currently the most studied, any mental state is represented by a considerable number of neuronal networks connected together (without forgetting glial cells!...). The present big projects like *Human brain project* will lead certainly to considerable progress such as new ways to cure neurodegenerative diseases or follow up of the evolution of a single neuron or, may be, creation of an artificial neuron. An enthusiasming story is now being written. And we can say the same remarks about **artificial intelligence**

**But on the strict point of view of pure scientific knowledge, will we be able to understand, to explain, why and how all these networks connect together?**

On this subject we propose, very unpretentiously, a way inspired by analogy with Quantum mechanics.

Conscious that we have only touched lightly these topics and that a considerable work remains to be done, we would be happy if scientists of different disciplines, more competent than us and interested by our hypothesis, could study thoroughly these subjects so as to elaborate a consistent theory.

# REFERENCES BIBLIOGRAPHY